\documentclass[aps,prl,graphicx,twocolumn,a4paper,superscriptaddress]{revtex4-1}

\usepackage{amsmath}    
\usepackage{graphicx}    
\usepackage{hyperref}   
\usepackage[english]{babel}

\usepackage[utf8]{inputenc}

\newcommand{\paral}{\uparrow\!\uparrow}
\newcommand{\antipar}{\uparrow\!\downarrow}

\newcommand{\quant}[2]{$#1\,\text{#2}$}

\begin{document}
\title{Electron-electron interaction strength in ferromagnetic
nickel determined by spin-polarized positron annihilation}
\author{Hubert Ceeh}
\affiliation{Technische Universit\"at M\"unchen, Lehrstuhl E21, James-Franck Stra\ss e, 85748 Garching, Germany}
\author{Josef-Andreas Weber}
\affiliation{Technische Universit\"at M\"unchen, Lehrstuhl E21, James-Franck Stra\ss e, 85748 Garching, Germany}
\author{Peter B\"oni}
\affiliation{Technische Universit\"at M\"unchen, Lehrstuhl E21, James-Franck Stra\ss e, 85748 Garching, Germany}
\author{Michael Leitner}
\affiliation{Heinz Maier-Leibnitz Zentrum (MLZ), Technische Universit\"at M\"unchen, Lichtenbergstra\ss e 1, 85748 Garching, Germany}
\author{Diana Benea}
\affiliation{Chemistry Department, University Munich, Butenandstra\ss e 5-13, 81377 M\"unchen, Germany}
\affiliation{Faculty of Physics, Babes-Bolyai University, Kog\u alniceanustr 1, 400084 Cluj-Napoca, Romania}
\author{Liviu Chioncel}
\affiliation{Augsburg Center for Innovative Technologies, University of Augsburg, D-86135 Augsburg, Germany}
\affiliation{Theoretical Physics III, Center for Electronic Correlations and Magnetism, Institute of Physics, University of Augsburg, 86135 Augsburg, Germany}
\author{Dieter Vollhardt}
\affiliation{Theoretical Physics III, Center for Electronic Correlations and Magnetism, Institute of Physics, University of Augsburg, 86135 Augsburg, Germany}

\author{Hubert Ebert}
\affiliation{Chemistry Department, University Munich, Butenandstra\ss e 5-13, 81377 M\"unchen, Germany}
\author{Jan Min\'ar}
\affiliation{Chemistry Department, University Munich, Butenandstra\ss e 5-13, 81377 M\"unchen, Germany}
\affiliation{New Technologies - Research Center, University of West Bohemia, Univerzitni 8, 306 14 Pilsen, Czech Republic}
\author{Christoph Hugenschmidt}
\affiliation{Technische Universit\"at M\"unchen, Lehrstuhl E21, James-Franck Stra\ss e, 85748 Garching, Germany}
\affiliation{Heinz Maier-Leibnitz Zentrum (MLZ), Technische Universit\"at M\"unchen, Lichtenbergstra\ss e 1, 85748 Garching, Germany}
\date{\today}
\begin{abstract}
The two-photon momentum distribution of annihilating
electron-positron pairs in ferromagnetic nickel (Ni) was determined by measuring the spin-polarized
two-dimensional angular correlation of annihilation radiation (ACAR). The spectra were compared with theoretical results obtained within LDA+DMFT, a combination of the local density approximation (LDA) and the many-body dynamical mean-field theory (DMFT). The self-energy describing the electronic correlations in Ni is found to make important anisotropic contributions to the momentum distribution which are not present in LDA. Based on a detailed comparison of the theoretical and experimental results the strength of the local electronic interaction $U$ in ferromagnetic Ni is determined as  $2.0\pm 0.1$\,eV.
\end{abstract}

\maketitle

\hspace*{-\parindent}
The electronic properties of many narrow-band materials, such as the $d$-shell transition-metal series and their compounds, cannot be explained within a one-electron picture, because there exist strong correlations between electrons in the partially filled $d$ band
\cite{Imada98,georges96,KV04,Kotliar06}.
Such systems are therefore better described by
multi-band models such as the Hubbard or Anderson-type lattice model. In these models  the \emph{local} Coulomb repulsion  $U$ is assumed to be the dominant interaction between the electrons. The ``Hubbard'' parameter $U$ was originally
introduced for single-band models~\cite{hubb.63,gutz.63} and is defined as the Coulomb-energy required
to place two electrons on the same site: $U=E(d^{n+1})+E(d^{n-1})-2E(d^{n})$. Here
$E(d^n)$ represents the total energy of a system for which $n$ electrons fill
a given $d$-shell on a given atom. In multi-band systems $U$
takes the form of an interaction matrix.

The Hubbard model and related lattice models are able to explain important general
features of correlated electrons, but they cannot describe the physics of real
materials in any detail. Namely, for an approach to be realistic it must take into account the explicit electronic and lattice structure of the systems.
Here the LDA+DMFT approach has led to great progress in the understanding of correlated electron materials~\cite{KV04,ko.sa.06,held.07,Anisimov97a,Licht+Kats97,psi-k}.
LDA+DMFT  is a computational scheme where
the local density
approximation (LDA) or the generalized-gradient approximation (GGA) provide
the material dependent input (orbitals and hopping parameters), while DMFT~\cite{MV89,georges96} 
solves the interacting local many-body problem originating from the local Hubbard interaction $U$ and Hund's rule coupling $J$.
The results can be compared with experimental data obtained, for example, by photoemission spectroscopy. In particular, this technique measures spectral functions, i.e., the imaginary part of a one-particle Green function, and thus determines correlation induced shifts of the spectral weight. This allows one to estimate the local Hubbard interaction $U$ of a material, say, Ni.
Indeed, most
investigations on the electronic structure of Ni relied on photoemission
spectroscopy~\cite{Aebi96,Schneider96}. Braun \emph{et al.}~\cite{Braun12}
demonstrated  the importance of local correlations in Ni by exploiting the magnetic
circular dichroism in bulk sensitive soft X-ray photoemission measurements. One of the dominant correlation effects observed in the photoemission
data for Ni is the satellite peak
situated at $6\,$eV below the Fermi level  \cite{SBM+12,Himpsel79}. This feature is not captured by LDA, but is well explained by LDA+DMFT ~\cite{Lichtenstein01}.
LDA+DMFT also reproduces the correct width of
the occupied  $3d$ bands and the exchange splitting~\cite{Lichtenstein01}.
A fundamental difficulty concerning the interpretation of photoemission data is the fact that
they
involve an interaction
of photons with matter.
This makes the determination of a parameter such as the Hubbard $U$ quite difficult. Namely, it is not only necessary to describe the excitation process, but also the propagation of the photoelectrons in the material as well as the process of detection itself. Therefore the experimental data, and the values of the interaction parameters derived from them, will be strongly influenced by the surface of a sample.

\begin{figure}[h]
\includegraphics[width=0.5\textwidth]{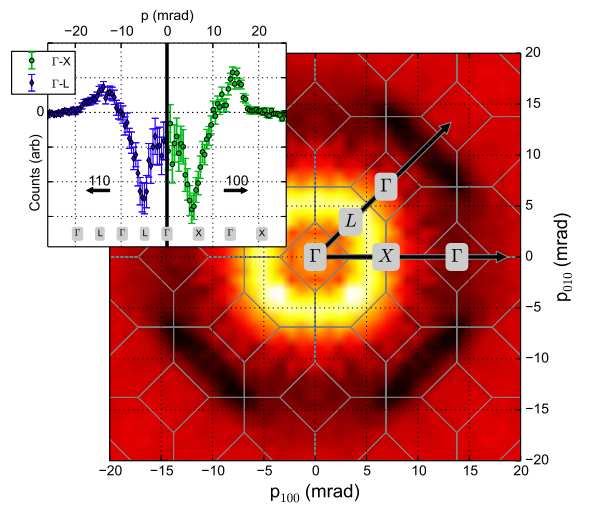}
\caption{\label{fig:Diff}(Color Online)
2D-ACAR difference spectra $\Delta N_{exp}(p_x,p_y)$ obtained when
the magnetic field is reversed, with the integration along the
$\langle001\rangle$ direction, $p_x=\langle100\rangle$ and
$p_y=\langle010\rangle$. The inset illustrate the anisotropy of the
difference spectra between two directions in momentum space.}
\end{figure}

In this Letter we discuss an alternative experimental technique
to determine the local Coulomb parameter $U$, involving \emph{positrons}.
In contrast to photoemission experiments
positron spectroscopy experiments
measure a two-particle Green function. Since there
are no physical processes except the electron-positron annihilation, positron
spectroscopy does not suffer from the above-mentioned difficulties faced by photoelectron
spectroscopy. As a consequence the magnitude of $U$ deduced
from positron spectroscopy is much less influenced by external effects (e.g., surfaces), which are difficult to control.
Here we show that by combining experimental results
of the spin-polarized two-dimensional angular correlation of annihilation radiation (2D-ACAR) with LDA+DMFT computations it is possible
to assess the strength of the electronic interactions in Ni quite unambiguously.
\begin{figure}[h]
\includegraphics[width=0.4\textwidth,clip=true]{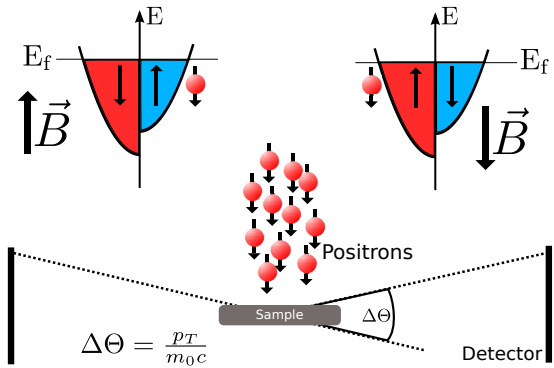}
\caption{(Color Online) \label{fig:prinzip} Schematic principle of spin polarized
2D-ACAR: In electron-positron annihilation the singlet configuration  is preferred 
for majority or minority spin electrons if the magnetization
of the sample is parallel or anti-parallel to the emission direction of
the positrons.}
\end{figure}
We measured spin-resolved two-dimensional projections of the
two-photon momentum distribution (TPMD)
in magnetic fields up to  $1.1\,$T at room temperature. The
field was applied parallel and anti-parallel, respectively,
relative to the crystallographic
$\langle110\rangle$ orientation of the sample which coincides with the emission
direction of the positrons (i.e., the polarization direction of the positron
beam). The $\beta^+$ spectrum of $^{22}$Na, which served as positron source,
has an end-point energy of $545\,$keV with a mean energy of about $215\,$keV.
This corresponds to a beam polarization of $74\,\%$ for forward emission.
Both, the positron back reflection in the source capsule and the magnetic guiding field, lower the net polarization of the beam, which was determined as
$34.4(5)\,\%$ in a separate experiment.
Although the positron undergoes multiple scattering during
thermalization~\cite{Zitzewitz79,Kubica75,Seeger87}
the polarization of the positrons remains essentially unchanged
until the moment of annihilation.
Quantum electrodynamics predicts the annihilation rate
in the triplet configuration to be significantly smaller ($1/1115$)
than that in the singlet configuration~\cite{ore49,Berko71}.
When an external magnetic field is applied parallel $(\paral)$ or anti-parallel $(\antipar)$
to the emission direction the positrons will therefore
annihilate  predominantly
with
electrons from the majority or the minority spin directions, respectively (see
Fig.~\ref{fig:prinzip}).
The result of a magnetic 2D-ACAR measurement can be expressed as~\cite{Berko71}
\begin{equation}\label{dNexp}
\Delta N_\text{exp}(p_x,p_y) = N_{\paral}(p_x,p_y)- N_{\antipar}(p_x,p_y)
\end{equation}
where $N_{\paral (\antipar)}(p_x,p_y)$ is the number of coincident photon counts measured with the
positron spin aligned parallel $(\paral)$ and anti-parallel  $(\antipar)$, respectively,  to the
applied field. For a detailed description of the experimental technique we
refer to Refs.~\cite{West81,Ceeh13,Leitner12}. Spin-polarized 2D-ACAR is one of
the few experimental methods that can probe the momentum distribution of the
electrons in the bulk with respect to the spin direction. It was successfully
applied to elemental Ni~\cite{Gen91} and other materials
\cite{Mijn86,Mijn90,Livesay01,Dug12}.
In Fig.~\ref{fig:Diff} we present the spin-difference of the 2D-ACAR
measurement of Ni. In each spectrum an excess of 250 million counts was
collected, and the data were corrected for the momentum sampling function.
Before subtraction, the spectra $N_{\paral}$ and $N_{\antipar}$ were normalized to an equal
amount of counts. A renormalization due to 3$\gamma$ decay was omitted since
the corresponding correction in the case of Ni \cite{Berko71} is negligible
compared to the statistical noise.
The 4-fold symmetry is clearly observed in agreement with the study of
Manuel \textit{et al.}~\cite{Gen91}. It should be noted that the difference
spectrum (see inset of Fig.~\ref{fig:Diff}) is anisotropic, i.e., the signal 
is more intense along the $\Gamma-X-\Gamma$ direction
than along the $\Gamma-L-\Gamma$ direction.

To represent the difference spectra $\Delta N_{exp}(p_x,p_y)$ in \textbf{k}-space we
apply the magnetic Lock-Crisp-West (LCW) back-folding procedure. In practice
this amounts to taking the difference of the individual spin-resolved
LCW-projections~\cite{LCW} since the LCW-procedure is nothing but a linear
transformation. According to this scheme, first proposed by
Biasini \textit{et al.}~\cite{Biasini06,Biasini07}, effects due to the
electron-positron interaction, known as enhancement, are cancelled out. This proves to be particularly advantageous as  no complete microscopic description for  the enhancement problem is available. Several recipes  have been discussed in the literature~\cite{JarlSigh85,JarlSingh87,Niem86,Dug10,Makkonen14,Sznaid14,Sznaid12} and have also been consequently applied for Ni~\cite{Sing86,Sznaid75}
Hence, the only parameter in this analysis is a uniform scale factor, which is
obtained by fitting the amplitude of theoretical spectra to experimental
data.

The theoretical analysis requires the knowledge of the two-particle
electron-positron Green function, describing the probability amplitude
for an electron and
a positron propagating between two different space-time points.
In view of the fact that experiments performed on
Ni~\cite{Mijnarends79,Barb97,Sznaid75} show no significant enhancement effects,
we factorize the two-particle Green function into a product of electronic
and positronic Green functions. Thereby Coulomb interaction induced correlation effects between the 
annihilating particles are neglected.  At the same time correlations between the \emph{electrons} are explicitly included in the DMFT.

Electronic structure calculations were performed with the
spin-polarized relativistic Korringa-Kohn-Rostoker (SPR-KKR) method~\cite{EKM11}.
For LDA computations the exchange-correlation potentials parametrized by Vosko, Wilk
and Nusair~\cite{VWN80} were used with a lattice
parameter of $3.52\, \AA$. 
To include the electronic correlations, a charge and self-energy
self-consistent LDA+DMFT scheme was employed, which is based on the KKR approach~\cite{mi.ch.05} and where the impurity problem is solved with a spin-polarized
$T$-matrix fluctuation exchange
(SPTF) method~\cite{Licht+Kats97,PKL05}. 
This impurity solver is fully rotationally invariant even in the multi-orbital version and  is reliable when the
interaction $U$ is smaller than the bandwidth, a condition which is fulfilled in the case of Ni.
In this LDA+DMFT framework
the electron-positron momentum density $\rho_{\sigma}({\bf p})$ is computed
directly from the two-particle Green function in the momentum representation.
The neglect of electron-positron correlations corresponds to the factorization
of the two-particle Green function in real space.
In the numerical implementation the position-space integrals for
the ``auxiliary'' Green function $G_{\sigma \sigma^{\prime}}({\bf p}_e,{\bf p}_p)$ obtained within LDA or LDA+DMFT, respectively,
are performed as integrals
over unit cells:
\begin{equation}
\begin{split}
&G^{\alpha}_{\sigma \sigma^{\prime}}({\bf p}_e,{\bf p}_p, E_e, E_p) =
\frac{1}{N \Omega}\int d^3{\bf r} \int d^3{\bf r}^{\prime}\\
  &\phi_{{\bf p}_e \sigma}^{e\dagger}({\bf r}) \,
Im \, G^{\alpha}_{e+}({\bf r},{\bf r}^{\prime},E_e) \,
\phi_{{\bf p}_e \sigma}^{e}({\bf r}^{\prime}) \\
  &\phi_{{\bf p}_p \sigma^{\prime}}^{p\dagger}({\bf r}) \,
Im \, G_{p+}({\bf r}, {\bf r}^{\prime},E_p) \,
\phi_{{\bf p}_p \sigma^{\prime}}^{p}({\bf r}\,')\;,
\end{split}
\end{equation}
where $\alpha$ = LDA or LDA+DMFT, and $({\bf p}_e, \sigma)$, and
$({\bf p}_p,\sigma^{\prime})$ are the momenta and spin of electron and
positron, respectively. Here $G^{\alpha}_{\sigma \sigma^{\prime}}$ is
computed for each energy point on the complex energy contour, providing the
electron-positron momentum density:
\begin{equation} \label{rho_ep}
 \rho_{\sigma}^{\alpha}({\bf p}) =  -\frac{1}{\pi} \int dE_e
G^{\alpha}_{\sigma \sigma^{\prime}}({\bf p}_e,{\bf p}_p, E_e, E_p).
\end{equation}
Integration over positron energies $E_p$ is not required, since only the
ground state is considered, and $\sigma^\prime = -\sigma$ at the
annihilation. The momentum carried off by the photons is equal
to that of the two particles up to a reciprocal lattice vector, reflecting
the fact that the annihilation takes place in a crystal. Hence an electron
with wave vector ${\bf k}$ contributes to
$\rho^{\alpha}_{\sigma}(\textbf{p})$ not
only at ${\bf p} = {\bf k}$ (normal process) but also at
${\bf p} = {\bf k} + {\bf K}$, with ${\bf K}$ a vector of the reciprocal
lattice (Umklapp process). The experimental spin-difference spectra
$\Delta N_{exp}(p_x,p_y)$ can be compared with the computed difference in
the integrated momentum densities of Eq.~\ref{rho_ep}:
\begin{equation}\label{dNtheo}
\Delta N_{theo}^{\alpha}(p_x,p_y) = \int
d p_z \left[ \rho_{\uparrow}^{\alpha}(\textbf{p}) -
\rho_{\downarrow}^{\alpha}(\textbf{p}) \right].
\end{equation}

\begin{figure}[th]
\includegraphics[width=0.5\textwidth]{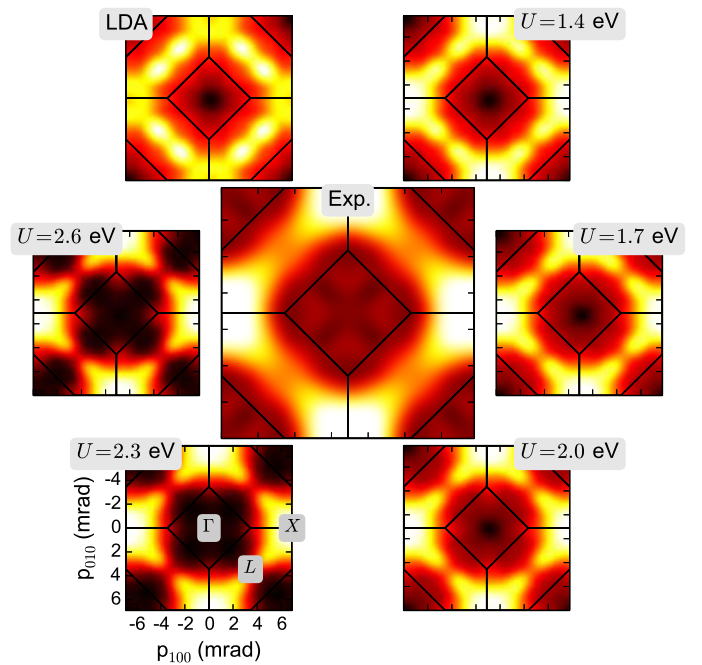}
\caption{(Color Online) \label{fig:lcw-diff-vgl}Experimental magnetic LCW
spectrum (\textit{center}) compared to theoretical spectra computed for
different values of the local Coulomb parameter $U$ in the range from
$1.4$ to \quant{2.3}{eV}.
}
\end{figure}

In Fig. ~\ref{fig:lcw-diff-vgl} we show the measured $\Delta N_{exp}(p_x,p_y)$
and the theoretical $\Delta N_{theo}^{\alpha}(p_x,p_y)$ LCW-folded
difference spectra for different values of $U$.
%
It is clearly seen that with increasing $U$ a gap opens at the L-points
of the Brillouin zone. This gap is associated with the necks in the FS of Ni.
Apparently LDA underestimates the density at the X-point, while the
density near the L-point is overestimated. In the LDA+DMFT calculation the
highest density is found at the X-point, similar to the experimental data.
However, the structure connecting the X and L-points is less 
pronounced in the experimental data than in the LDA+DMFT results.
Obviously the local interaction (see e.g.~\cite{Braun12}) pushes the $d$-bands below the Fermi energy, whereby the X$_2$ hole pocket obtained in LDA disappears. This also greatly changes the calculated anomalous Hall effect of Ni \cite{PhysRevB.84.144427,KCME13}.
\begin{figure}[h]
\includegraphics[width=0.5\textwidth,clip=true]{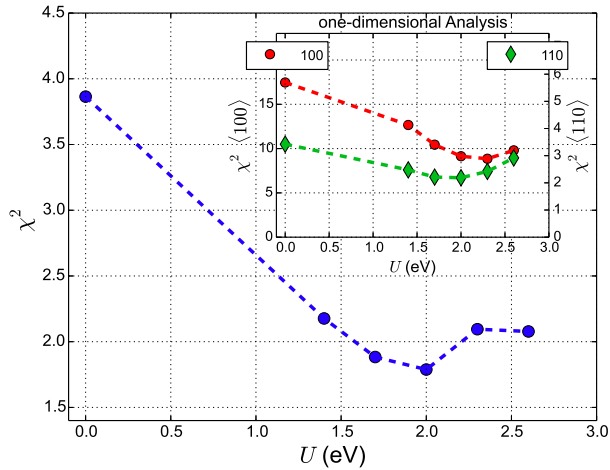}
\caption{(Color Online) \label{fig:Chi2} Least square fit analysis ($\chi^2$) between LDA+DMFT calculations and experimental data as a function of the Hubbard $U$ for the 2D data. Higher $U$ values correspond to stronger electron-electron correlations. A pronounced minimum of $\chi^ 2$ is found for $U=2.0\,$eV. The inset shows the results for the 1D data. (The dotted lines act as a guide to the eye.) 
}
\end{figure}

In order to derive the value of the local Coulomb interaction parameter $U$
we performed a least square fit analysis of the measured data with the
LDA+DMFT calculations (Fig.~\ref{fig:lcw-diff-vgl}).
The result is summarized in Fig.~\ref{fig:Chi2}.
A distinct minimum in $\chi^2$ is found at $U=2.0\,$eV.
The 1D data are shown in the inset of Fig.~\ref{fig:Chi2}.
Although the spin difference spectra  along
different directions in momentum space are not the same, the $\chi^2$ curves have their minima 
in the same region around $U=2.0\,$eV of the Hubbard interaction. The loss of information in the twice-integrated 1D data is indicated by a larger $\chi^2$ value.
Interpolating the data in Fig.~\ref{fig:Chi2} with higher order
polynomials allows us to estimate the systematic error in the position of
the absolute minimum as $\pm 0.1\,$eV.
The resulting value $U=2.0 \pm 0.1\,$eV supports the conclusion of recent measurements  of the Compton scattering
profiles~\cite{ch.be.14a,ch.be.14b} where values of $2.0\,$eV and
$2.3\,$eV for $U$  were reported.

To conclude, we have shown that spin-polarized 2D-ACAR is a powerful tool
to investigate the electronic structure of ferromagnetic systems especially
when combined with LDA+DMFT calculations.  Here the strength of magnetic 2D-ACAR becomes apparent: the higher information content of the two-dimensional data compared to twice-integrated one-dimensional data from Compton scattering.
We found strong evidence for the existence of electron pockets around the
X point which are related to electronic correlation effects. A detailed
comparison of the experimental results with electronic structure
calculations in the framework of LDA+DMFT allowed us to evaluate the local
Coulomb interaction (``Hubbard'' $U$) as $2.0 \pm 0.1\,$eV.

This project was funded by the Deutsche Forschungsgesellschaft (DFG) through the Transregional Collaborative Research Center TRR 80 and Research Unit FOR1346. DB acknwledges the DAAD and the CNCS - UEFISCDI(PN-II-ID-PCE-2012-4-0470). JM acknowledge also the CENTEM project (CZ.1.05/2.1.00/03.0088 co-funded by the ERDF).

\bibliography{Ni2n}

\end{document}